\documentclass[pra,twocolumn,superscriptaddress,floatfix,amsmath,amssymb,bibnotes]{revtex4-1}

\usepackage[english]{babel}
\usepackage{graphicx}
\usepackage{bm}
\usepackage{braket}
\usepackage{amsfonts}
\usepackage{comment}
\usepackage{hyperref}

\begin{document}


\title{Learning to learn with quantum neural networks via classical neural networks}

\affiliation{Google LLC, Venice, CA 90291
}

 \affiliation{Department of Applied Mathematics, University of Waterloo, Waterloo, Ontario, N2L 3G1, Canada
}
 \affiliation{Department of Computer Science, University of Waterloo, Waterloo, Ontario, N2L 3G1, Canada
}

 \affiliation{Institute for Quantum Computing, University of Waterloo, Waterloo, Ontario, N2L 3G1, Canada
}
\affiliation{Department of Electrical Engineering and Computer Science, University of Michigan, Ann Arbor, MI 48109}

\author{Guillaume Verdon}
\email{Both authors contributed equally to this work.}
\affiliation{Google LLC, Venice, CA 90291
}

 \affiliation{Department of Applied Mathematics, University of Waterloo, Waterloo, Ontario, N2L 3G1, Canada
}
 \affiliation{Institute for Quantum Computing, University of Waterloo, Waterloo, Ontario, N2L 3G1, Canada
}

\author{Michael Broughton}
\email{Both authors contributed equally to this work.}
\affiliation{Google LLC, Venice, CA 90291
}

 \affiliation{Department of Computer Science, University of Waterloo, Waterloo, Ontario, N2L 3G1, Canada
}

\author{Jarrod R. McClean}
\affiliation{Google LLC, Venice, CA 90291
}

\author{Kevin J. Sung}

\affiliation{Google LLC, Venice, CA 90291
}

\affiliation{Department of Electrical Engineering and Computer Science, University of Michigan, Ann Arbor, MI 48109}

\author{Ryan Babbush}
\affiliation{Google LLC, Venice, CA 90291
}
\author{Zhang Jiang}
\affiliation{Google LLC, Venice, CA 90291
}

\author{Hartmut Neven}
\affiliation{Google LLC, Venice, CA 90291
}

\author{Masoud Mohseni}
\affiliation{Google LLC, Venice, CA 90291
}

\date{\today}

\begin{abstract} 
Quantum Neural Networks (QNNs) are a promising variational learning paradigm with applications to near-term quantum processors, however they still face some significant challenges. One such challenge is finding good parameter initialization heuristics that ensure rapid and consistent convergence to local minima of the parameterized quantum circuit landscape. In this work, we train classical neural networks to assist in the quantum learning process, also know as meta-learning, to rapidly find approximate optima in the parameter landscape for several classes of quantum variational algorithms. Specifically, we train classical recurrent neural networks to find approximately optimal parameters within a small number of queries of the cost function for the Quantum Approximate Optimization Algorithm (QAOA) for MaxCut, QAOA for Sherrington-Kirkpatrick Ising model, and for a Variational Quantum Eigensolver for the Hubbard model. By initializing other optimizers at parameter values suggested by the classical neural network, we demonstrate a significant improvement in the total number of optimization iterations required to reach a given accuracy. We further demonstrate that the optimization strategies learned by the neural network generalize well across a range of problem instance sizes. This opens up the possibility of training on small, classically simulatable problem instances, in order to initialize larger, classically intractably simulatable problem instances on quantum devices, thereby significantly reducing the number of required quantum-classical optimization iterations.
\end{abstract}

\maketitle
 

\section{Introduction}
With the advent of noisy intermediate-scale quantum (NISQ) devices \cite{preskill2018quantum}, there has been a growing body of work \cite{farhi2018classification,farhi2014quantum,peruzzo2014variational,killoran2018continuous,wecker2015progress,biamonte2017quantum,zhou2018quantum,mcclean2016theory,hadfield2017quantum,grant2018hierarchical,khatri2019quantum,schuld2019quantum,mcardle2018variational,benedetti2019adversarial,nash2019quantum,jiang2018majorana,steinbrecher2018quantum,fingerhuth2018quantum,larose2018variational,cincio2018learning,situ2019variational,chen2018universal,verdon2017quantum,preskill2018quantum} 
aiming to develop algorithms which are suitable to be run in this near-term era of quantum computing. A particularly promising category of such algorithms are the so-called quantum-classical variational algorithms \cite{peruzzo2014variational,mcclean2016theory}, which involve the optimization over a family of parameterized quantum circuits using classical optimization techniques (see Fig. \ref{fig:q-c-graph}). These variational algorithms are promising because they have flexible architectures, are adaptive in nature, can be tailored to fit the gate allowances of near-term quantum devices, and are partially robust to systematic noise. 

Many quantum-classical variational algorithms consist of optimizing the parameters of a parameterized quantum circuit to extremize a cost function (often consisting of the expectation value of a certain observable at the output of the circuit). This optimization of parameterized functions is similar to the methods of classical deep learning with neural networks \cite{lecun2015deep,goodfellow2016deep,schmidhuber2015deep}. Furthermore, the training and inference processes for classical deep neural networks have been shown to be embeddable into this quantum-classical PQC optimization framework \cite{verdon2018universal,killoran2018continuous}. Given these connections, it has become common to sometimes refer to certain PQC ansatze as \textit{Quantum Neural Networks} \cite{farhi2018classification,chen2018universal,mcclean_boixo_smelyanskiy_babbush_neven_2018,biamonte2017quantum} (QNN's). 

Optimization of QNN's in the NISQ era is currently faced with two main challenges. The first is local optimization; the stochastic nature of the objective function in combination with readout complexity considerations has made direct translation of classical local optimization algorithms challenging. Proposed gradient-based optimizers either rely on a quantum form of backpropagation of errors \cite{verdon2018universal} that requires additional gate depth and quantum memory, or use finite-difference gradients \cite{farhi2018classification,chen2018universal} which typically require numerous quantum circuit evaluations for each gradient descent iteration. Recent works have proposed sampling analytic gradients \cite{schuld2018evaluating,
harrow2019low} to reduce this cost. However, these approaches also require many measurement runs and consequently remain expensive, and further advances are needed in this area. 

The second major challenge for QNN optimization is parameter initialization. Although there have been some proposals for QNN parameter initialization heuristics \cite{yang2017optimizing,zhou2018quantum,grant2019initialization}, we believe there is a need for more efficient and more flexible variants of such heuristics. By initializing parameters in the neighborhood of a local minimum of the cost landscape, one ensures more consistent local optimization convergence in a fewer number of iterations and a better overall answer with respect to the global landscape. Good initialization is thus crucial to promote the convergence of local optimizers to local extrema and to select reasonably good local minima.

In this paper, we tackle the second problem of parameter initialization by exploring methods that leverage classical neural networks trained to optimize control parameters of parametrized quantum circuits. Taking inspiration from the growing body of work on meta-learning, also known as \textit{learning to learn} \cite{chen2016learning,andrychowicz2016learning,zoph2016neural,finn2017model,long2015learning}, we use a classical recurrent neural network (RNN) as a black-box controller to optimize the PQC parameters directly, as shown in Figure \ref{fig:RNN}. We train this RNN using random problem instances from specific classes of problems. We explore the performance of this approach for the following problem classes: quantum approximate optimization algorithm (QAOA) for MaxCut \cite{farhi2014quantum}, QAOA for Sherrington-Kirkpatrick Ising models \cite{yang2017optimizing}, and a Variational Quantum Eigensolver (VQE) ansatz for the Hubbard model \cite{kivlichan2018quantum,jiang2018quantum,wecker2015progress}. 

Through numerical simulations, we show that a recurrent neural network trained to optimize small quantum neural networks can learn parameter update heuristics that generalize to larger system sizes and problem instances, while still outperforming other initialization strategies at this scale. This opens up the possibility of classically training RNN optimizers for specific problem classes using instances of classically simulatable QNN's with reasonable system sizes as training data. After this training is done, these RNN optimizers could then be used on problem instances with QNN's whose system sizes are beyond the classically simulatable regime. 

For reasons explained further in sections \ref{sec:Results}, we use the RNN as a few-shot global approximate optimizer, which is used to initialize a local optimizer, such as Nelder-Mead \cite{lagarias1998convergence,nannicini2019performance,guerreschi2017practical}. In principle, the neural network could initialize any other local optimizer (such as SPSA \cite{spall1998implementation}, BOBYQA \cite{powell2009bobyqa}, and many others \cite{nannicini2018performance}), however, the focus of this paper is not to benchmark and compare various options for local optimizers. We have found that our approach compares favorably over other standard parameter initialization methods for all local optimizers studied. To the authors' knowledge, this work is a first instance where meta-learning techniques have been successfully applied to enhance quantum machine learning algorithms.

\section{Quantum-Classical Meta-Learning}
\subsection{Variational Quantum Algorithms}\label{sec:vqa}

\begin{figure} 
    \includegraphics[width=0.5\textwidth]{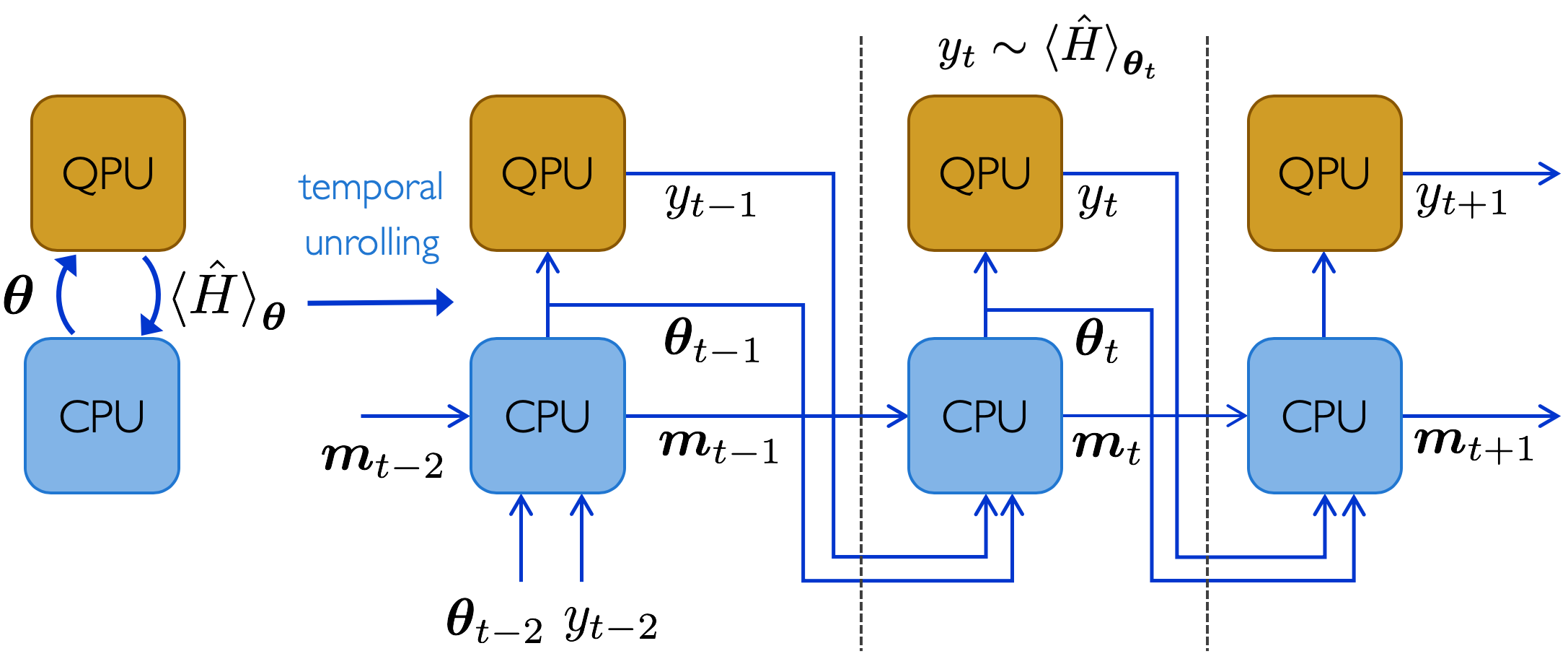}
    \caption{Unrolling the temporal quantum-classical hybrid computational graph of a general hybrid variational quantum algorithm. At the $t^\text{th}$ optimization iteration, the CPU is fed the previous iterations' parameters $\bm{\theta}_{t}$, and the expectation value of the Hamiltonian at the previous step $y_{t} = \braket{\hat{H}}_{\bm{\theta}_{t}}$, it also has access to its own internal memory $\bm{m}_t$. A classical optimization algorithm then suggest a new set of parameters $\bm{\theta}_{t}$, which is fed to the QPU. The QPU then executes multiple runs to obtain $y_{t+1}$, the expectation value of the cost Hamiltonian at the output of the parameterized quantum circuit evaluated at these given parameters.}
    \label{fig:q-c-graph}
\end{figure}

Let us first briefly review the theory of variational quantum algorithms, and how one can view the hybrid quantum-classical optimization process as a hybrid computational graph. Variational quantum algorithms are comprised of an iterative quantum-classical optimization loop between a classical processing unit (CPU) and a quantum processing unit (QPU), pictured in Figure \ref{fig:q-c-graph}. 

An iteration begins with the CPU sending the set of candidate parameters $\bm{\theta}$ to the QPU. The QPU then executes a parameterized circuit \(\hat{U}(\bm{\theta})\), which outputs a state \(\ket{\psi_{\bm{\theta}}} \). For the types of QNN of interest in this work, namely QAOA \cite{farhi2014quantum} and VQE \cite{peruzzo2014variational}, the function to be optimized is the expectation value of a certain Hamiltonian operator \(f(\bm{\theta})   \equiv \bra{\psi_{\bm{\theta}}}\hat{H}\ket{\psi_{\bm{\theta}}}\). We will refer to this function to as the \textit{cost function} of the variational quantum algorithm, defined by the \textit{cost Hamiltonian}. The expectation value of the cost Hamiltonian \(\braket{\hat{H}}_{\bm{\theta}} \equiv \bra{\psi_{\bm{\theta}}}\hat{H}\ket{\psi_{\bm{\theta}}}\) is estimated using the quantum expectation estimation procedure \cite{43965} via many repeated runs of the QPU. Following this, the estimated expectation value is relayed back to the CPU, where the classical optimizer running on the CPU is then tasked with suggesting a new set of parameters for the subsequent iteration \footnote{In general, one may relay the raw measurement results to the classical processing unit, which can then compute the expectation value of the cost function. For the purposes of this paper, we assumed the classical optimizer only has access to (noisy) estimates of the expectation value of the cost Hamiltonian.}. 

From an optimization perspective, the CPU is given a parametrized black-box function $f:\mathbb{R}^m\rightarrow \mathbb{R}$ for which it is tasked to find a set of parameters minimizing this cost function \(\bm{\theta}^\ast=  \text{arg}\!\min_{\bm{\theta}\in \mathbb{R}^m} f(\bm{\theta})\). In many cases, finding an approximate minimum is sufficient. Typically, one must consider this function to have a stochastic output which serves as a noisy unbiased estimate (under some assumptions) of the true output value of the function we are ultimately trying to optimize. Optimizing this output rapidly and accurately, despite only having access to noisy estimates poses a significant challenge for variational quantum algorithms.

Even for perfect quantum gates and operations, for a finite number of measurement runs, there is inherent noise in the quantum expectation estimate \cite{43965}. Usually when performing quantum expectation estimation, the cost Hamiltonian can be expressed as a linear combination of $k$-local Pauli's, 
\(
    \hat{H} = \sum_{j=1}^N \alpha_j \hat{P}_j,
\)
where the $\alpha_j$'s are real-valued coefficients and the $\hat{P}_j$'s are Paulis that are at most $k$-local \cite{kitaev2002classical}. The measurement of expectations of $k$-local Pauli observables is fairly straightforward \cite{mcclean2016theory}, while the linear combination of expectation values is done on the classical device. For a Hamiltonian $\hat{H}$ with such a decomposition, we define its \emph{Pauli coefficient norm}, denoted $\lVert \hat{H}\rVert_*$, as the one-norm of the vector of coefficients in its Pauli decomposition, namely  $\lVert \hat{H}\rVert_* \equiv \lVert\alpha\rVert_1 =  \sum_{j=1}^N |\alpha_j|$. For such Hamiltonians, the expected number of repetitions is bounded by $\sim\mathcal{O}(\lVert \hat{H}\rVert_*^2/\epsilon^2)$ to get an estimate that is accurate within $\epsilon$ from the unbiased value with a desired probability \cite{rubin2018application,wecker2015progress}.

Finally, note that the quantum-classical optimization loop can be unrolled over time into a single temporal quantum-classical computational graph, as depicted in Figure \ref{fig:q-c-graph}. This hybrid computational graph can be considered as hybrid quantum-classical neural networks. We developed methods to propagate gradients through such hybrid computational graph using reverse-mode auto-differentiation, also known as \textit{backpropagation} \cite{lecun1988theoretical,rumelhart1986learning}. We achieve this by converting hybrid quantum-classical backpropagation methods from previous work \cite{verdon2018universal}, originally formulated for quantum optimizers, to a form suitable for classical optimizers, which are most relevant for the NISQ era and are our focus in this paper. During the writing of this paper, other works have also employed hybrid quantum-classical backpropagation for various quantum machine learning tasks \cite{schuld2018evaluating,romero2019variational}.

\subsection{Meta-learning with Neural Optimizers} \label{sec:meta-l}

Meta-learning, also called \textit{learning to learn} \cite{andrychowicz2016learning,chen2016learning}, consists of a set of meta-optimization techniques which aim to learn how to modify the parameters (or hyperparameters) of learning algorithms to further tailor them for a specific purpose. This could be to ensure that the learning generalizes well (minimizes test set error), to better fit the given data in less iterations (minimize training set error) \cite{andrychowicz2016learning}, or to perform transfer learning (adapt a pre-trained neural network to a new task) \cite{finn2017model,nichol2018reptile}. In recent years, there have been many new works in the meta-learning literature \cite{andrychowicz2016learning,chen2016learning,finn2017model,nichol2018reptile,rusu2018meta}, and we aim to transfer some of the tools developed in the context of classical deep learning to quantum variational algorithms.

Our aim will be to train a classical \textit{optimizer} neural network to learn parameter update heuristics for \textit{optimizee} quantum neural networks. As mentioned previously, for our QNN's of interest, the cost function to be optimized is the expectation value of a certain Hamiltonian operator $\hat{H}$, with respect to a parameterized class of states $\ket{\psi_{\bm{\theta}}}$ output by a family of parametrized quantum circuits; $f(\bm{\theta}) = \braket{\hat{H}}_{\bm{\theta}}$. To emulate the statistical noise of quantum expectation estimation, we will this assume that the optimizer is fed noisy unbiased estimates of the QNN cost function at test time.

In many studies of meta-learning \cite{andrychowicz2016learning}, 
it is assumed that this black box (in our case a QNN) is differentiable and that the learner has oracular access to the gradients of the function with respect to its parameters, $\nabla f(\bm{\theta})$. Since precise gradient estimations on NISQ devices are hampered by the large number of runs required and by the noise of the device, we focus on the case where the learner will only have access to black box function queries at test time. We will, however, use gradients for neural optimizer network training. This access to gradients during training is not strictly necessary, but can speed up training in some cases. Note that gradients of hybrid quantum-classical computational graphs can be obtained by backpropagation (automatic differentiation) when simulating quantum circuits, or by using techniques for backpropagation through black boxes \cite{audet2016blackbox}, or hybrid quantum-classical backpropagation \cite{verdon2018universal,schuld2018evaluating}.

\begin{figure}
    \centering
    \includegraphics[width=0.44\textwidth]{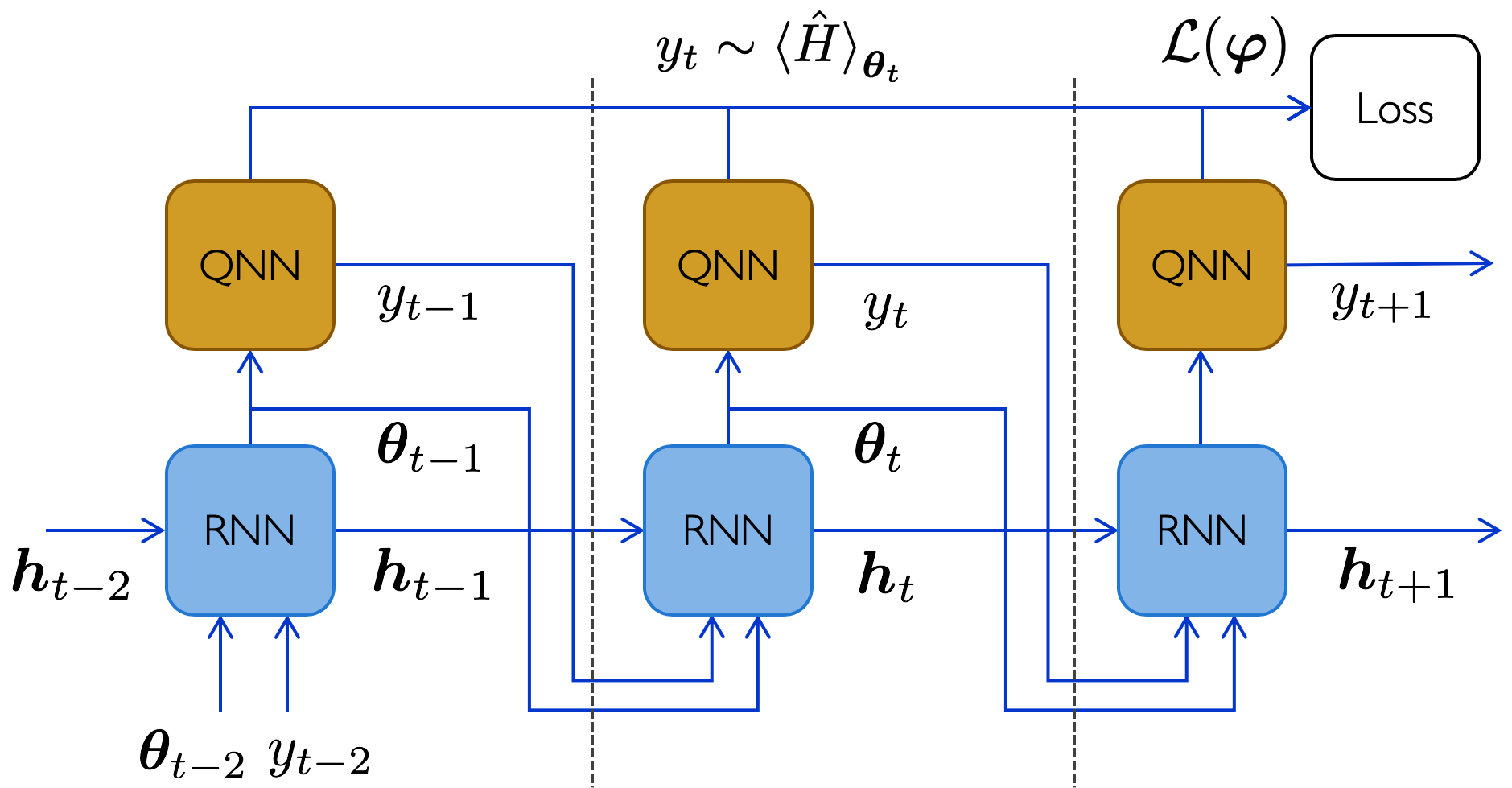}
    \caption{Unrolled temporal quantum-classical computational graph for the meta-learning optimization of the recurrent neural network (RNN) optimizer and a quantum neural network (QNN). This graph is similar to the general VQA graph in Figure \ref{fig:q-c-graph}, except that the memory of the optimizer is encoded in the hidden state of the RNN $\bm{h}$, and we represent the flow of data used to evaluate the meta-learning loss function. This meta loss function $\mathcal{L}$ is a functional of the history of expectation value estimate samples $\bm{y} = \{y_t\}_{t=1}^T$, and is thus indirectly dependent on the RNN parameters $\bm{\varphi}$. We see that backpropagating from the meta-loss node to the RNN's necessitates gradients to pass through the QNN. }
    \label{fig:RNN}
\end{figure}

To choose an architecture for the optimizer network, we interpret the QNN parameters and cost function evaluations over multiple quantum-classical iterations as a sequence-to-sequence learning problem. A canonical choice of neural network architecture for processing such sequential data is a recurrent neural network (RNN) \cite{lipton2015critical,pascanu2013difficulty}. Generally speaking, a recurrent neural network is a network where, for each item in a sequence, the network accepts an input vector, produces an output vector, and potentially keeps some data in memory for use with subsequent items. The computational graph of a RNN usually consists of many copies of the network, each sharing the same set of parameters, and each representing a time step. The  recurrent connections, which can be interpreted as self-connections representing the data flow over time, can be represented as a connections between copies of the network representing subsequent time steps. In this way, the computational graph can then be pictured in an \emph{unrolled} form, as depicted in Figure \ref{fig:RNN}. A particular type of RNN architecture which has had demonstrable successes over other RNN architectures is the Long Short Term Memory Network (LSTM) \cite{hochreiter1997long}. The LSTM owes its successes to its internal tunable mechanisms which, as its name implies, allow it to identify both long-term and short-term dependencies in the data. 

The meta-learning neural network architecture used in this paper is depicted in Figure \ref{fig:RNN}, there, an LSTM recurrent neural network is used to recursively propose updates to the QNN parameters, thereby acting as the classical black-box optimizer for quantum-classical optimization. At a given iteration, the RNN receives as input the previous QNN query's estimated cost function expectation $y_{t} \sim p(y|\bm{\theta}_t) $, where $y_t
$ is the estimate of $\braket{\hat{H}}_t$, as well as the parameters for which the QNN was evaluated $\bm{\theta}_t$. The RNN at this time step also receives information stored in its internal hidden state from the previous time step $\bm{h}_t$. The RNN itself has trainable parameters $\bm{\varphi}$, and hence it applies the parameterized mapping 
\begin{equation}
    \bm{h}_{t+1}, \bm{\theta}_{t+1} = \text{RNN}_{\bm{\varphi}}(\bm{h}_t,\bm{\theta}_{t},y_t)
\end{equation}
which generates a new suggestion for the QNN parameters as well as a new hidden state. Once this new set of QNN parameters is suggested, the RNN sends it to the QPU for evaluation and the loop continues. Note that this specific meta-learning architecture was adapted from previous work \cite{chen2016learning}, where the task of `learning to learn without gradient descent by gradient descent' was considered. 

The QNN architectures we chose to focus on were part of a class of ansatze known as Quantum Alternating Operator Ansatze \cite{hadfield2017quantum}, which are generalizations of the Quantum Approximate Optimization Algorithm \cite{farhi2014quantum}. These ansatze can be interpreted as a method for variationally-optimized bang-bang-controlled quantum-dynamical evolution in an energy landscape  \cite{farhi2014quantum,verdon2018universal,bapat2018bang,verdon2019quantum}. In this case, the QNN's variational parameters are the control parameters of the dynamics, and by appropriately tuning these parameters via quantum-classical optimization, one can cause the wavefunction to effectively descend the energetic landscape towards lower-energy regions. In recent works \cite{verdon2018universal,verdon2019quantum}, explicit connections between these quantum-dynamical control parameters and the hyperparameters of gradient descent algorithms were established for the continuum-embedded variants of QAOA. Thus, one can interpret these QAOA parameters as analogous to the hyperparameters of a quantum form of energetic landscape descent. 
In a recent work in classical meta-learning \cite{andrychowicz2016learning}, the authors use a RNN to control the hyperparameters of a neural network gradient-based training algorithm, drastically improving the neural networks training time and quality of fit when compared to stochastic gradient descent. It is thus natural to consider using an RNN to optimize QAOA parameters,  given the above intuition about their analogous relation to gradient descent hyperparameters. Another reason for choosing this particular type of QNN is that the number of variational parameters in these ansatze do not depend on the system size, only on the number of alternating operators. This means we can use this approach to train our RNN on smaller QNN instance sizes for certain problems, then test for generalization on larger QNN instance sizes.

Before we dive further into the specific quantum-classical meta-learning experiments, we will provide further details on how one would train an RNN to optimize such QNN's.

\subsubsection{Meta-Training \& Loss functions}

The objective of quantum-classical meta-learning is to train our RNN to learn an efficient parameter update scheme for a family of cost functions of interest, i.e., to discover an optimizer which efficiently optimizes a certain distribution of optimizees, on average. We consider an efficient optimizer to be one which finds sufficiently optimal approximate local minima of cost functions in as few function queries as possible. What qualifies as sufficiently optimal will depend on the class of problems at hand and the domain of application of interest. 

In the original work by DeepMind \cite{chen2016learning}, the neural optimizer was to be used as a general optimizer; little to no assumptions were made about the optimizee (the network being optimized) apart from the dimension of the parameter space.  The optimizer network was to be trained on one `data set' of optimizees, yet had to be applicable to a wide array of optimizees previously unseen. To learn such a general optimization strategy, the optimizer RNN was trained on random instances of a fairly general distribution of functions, namely, functions sampled from Gaussian processes \cite{rasmussen2004gaussian}.

Since we are focused on QNN optimization landscapes which are known to differ from classical Gaussian process optimization landscapes \cite{mcclean_boixo_smelyanskiy_babbush_neven_2018} 
we instead aimed to train specialized neural optimizers that are tailored to specific classes of problems and QNN ansatze. To explore how effective this is, we trained our RNN on random QNN instances within a targeted class of problems, namely QAOA and VQE, and tested the trained network on (larger) previously unseen instances from their respective classes. In Section \ref{sec:ansatze}, we describe in greater detail the various classes of problems and corresponding ansatze which were considered for training and testing.

Given a distribution of optimizees of interest, we must choose an adequate meta-learning loss function $\mathcal{L}(\bm{\varphi})$ with respect to which we will want to optimize the RNN parameters $\bm{\varphi}$. For a given QNN with cost function $f(\bm{\theta}) = \braket{\hat{H}}_{\bm{\theta}}$, we know that the RNN's meta-learning loss function \(\mathcal{L}(\bm{\varphi})\) will generally be dependent on the estimated cost function (quantum expectation estimate) history \(\{\mathbb{E}_{f,\bm{y}}[f({\bm{\theta}}_t)]\}_{t=1}^T\), but there is some flexibility in choosing exactly what this dependence is. Choosing the appropriate meta-learning loss function for the task at hand can be tricky, and depends on what is the particular application of the QNN. To be most general, we will want to pick a loss function which can learn to rapidly find optima of the parameter landscape yet is still constantly driven to find higher quality optima.

A simple choice of meta-learning loss function would be to use the expected final cost value at the end of the optimization time horizon, averaged over our samples from our distribution of functions $f$, i.e., $\mathcal{L}(\bm{\varphi}) = \mathbb{E}_{f,y_T}[f({\theta}_T)]$. In practice, this is a sparse signal, and would require backpropagation through a large portion of the computational graph for the loss signal to reach early portions of the RNN graph. A practical option of the same vein is the \textit{cumulative regret}, which is simply the sum of the cost function history uniformly averaged over the time horizon \(\mathcal{L}(\bm{\varphi}) = \sum_{t=1}^T \mathbb{E}_{f,\bm{y}}[f({\theta}_t)]\). This is a better choice as the loss signal is far less sparse, and the cumulative regret is a proxy for the minimum value achieved over the optimization history. In practice, this loss function may not be optimal as it will prioritize rapidly finding an approximate optimum and staying there. What is needed instead is a loss function that encourages exploration of the landscape in order to find a better optimum. The loss function we chose for our experiments is the \textit{observed improvement} at each time step, summed over the history of the optimization: 
\begin{equation}\label{eq:OI}
    \mathcal{L}(\bm{\varphi}) = \mathbb{E}_{f,\bm{y}}\left[\textstyle\sum\limits_{t=1}^T \text{min}\{f(\bm{\theta}_t) - \min_{j<t}[f(\bm{\theta}_j)], 0 \}\right],
\end{equation}
 The observed improvement at time step $t$ is given by the difference between the proposed value, $f(\bm{\theta}_t)$, and the best value obtained over the past history of the optimization until that point, $\min_{j<t}[f(\bm{\theta}_j)]$. If there is no improvement at a given time step then the contribution to the loss is nil. However, a temporary increase of the cost function followed by a significant improvement over the historical best will be rewarded rather than penalized (in contrast to the behavior of the cumulative regret loss).

In order to train the RNN, we need to differentiate the above loss function $\mathcal{L}(\bm{\varphi})$. One option to achieve this is via backpropagation of gradients through the unrolled RNN graph (depicted in Fig. \ref{fig:RNN}). This approach is called \textit{backpropagation through time}, and can be tricky to scale to arbitrarily deep networks due to vanishing/exploding gradient problems \cite{pascanu2013difficulty}. For practical purposes, a small time horizon is preferable, as it limits the complexity of the training of the RNN optimizer and avoids the pathologies of backpropagation through long time horizons. Since our loss function $\mathcal{L}(\bm{\varphi})$ is dependent on the QNN evaluated at multiple different parameter values \(\{\bm{\theta}_t\}_{t=1}^T\), in order to perform backpropagation through time we need to backpropagate gradients through multiple instances of the QNN.  

As our approach was backpropagation-based, to avoid problems of gradient blowup and to minimize the complexity of training, we keep a small time horizon for our numerical experiments featured in Section \ref{sec:Results}. As such, our RNN optimizer is intended to only run for a fixed number of iterations, and will be used as an initializer for other optimizers that perform local search. In principle, one could let the RNN optimize over more iterations at inference time than it was originally trained for, though the performance for later iterations may suffer. In our case the output of the RNN optimizer after a fixed number of iterations is used to initialize the parameters of the QNN's near a typical optimal set of parameters. It has been observed \cite{yang2017optimizing}, and in some cases formally proven \cite{brandao2018fixed}, that QAOA-like ansatze have a concentration of optimal parameters. Thus, the neural optimizer is used to learn a problem-class-specific initialization heuristic, and the fine-tuning is left for other optimizers. Since the neural optimizer would eventually learn a local heuristic for the fine-tuning, the added complexity cost of training for long time horizons if not justified by corresponding improvements in optimization efficiency, and we find that the combination of the neural optimizer as initializer and a greedy heuristic such as Nelder-Mead works quite well in practice, as shown in Figure \ref{fig:results}. Now, let us cover which ansatze we applied our RNN to learn to optimize.

\section{Numerical Experiments}
\label{sec:ansatze}

In this section, we provide a brief overview of the quantum neural network ansatze and problem instances considered for the hybrid meta-learning numerical experiments (results presented in Section \ref{sec:Results}). 
We trained and tested different `specialist' RNN optimizers for each of these three problem classes: quantum approximate optimization for MaxCut (MaxCut QAOA), quantum approximate optimization for Sherrington-Kirkpatrick Ising models \cite{yang2017optimizing} (Ising QAOA), and a Trotter-based variational quantum eigensolver ansatz for the Hubbard model \cite{kivlichan2018quantum} (Hubbard VQE).
We provide a brief introduction to each of these three classes, as well as describe the distribution of instances from these classes from which we sampled to generate training and testing instances.

\subsection{Quantum Approximate Optimization Algorithms}\label{sec:QAOA}

Let us first introduce a general QAOA ansatz before we specialize to applications to MaxCut problems and Ising (Sherrington-Kirkpatrick; SK) Hamiltonians. The goal of the QAOA is to  prepare low-energy states of a \textit{cost Hamiltonian} $\hat{H}_C$, which is usually a Hamiltonian which is diagonal in the computational basis. To achieve this, we typically begin in an eigenstate of a \textit{mixer Hamiltonian} $\hat{H}_M$, which does not commute with the cost Hamiltonian; $[\hat{H}_C,\hat{H}_M] \neq 0$. Applied onto this initial state is a sequence of exponentials of the form
\begin{equation}\label{eq:QAOA}
   \hat{U}(\bm{\theta}) = \prod_{j=1}^{P}e^{-i\theta^{(j)}_m\hat{H}_M}e^{-i\theta^{(j)}_{c} \hat{H}_C}, 
\end{equation}
where $\bm{\theta} = \{\bm{\theta}_m,\bm{\theta}_c\}$ are variational parameters to be optimized. Note that in the above and throughout this paper we will use the operator product notation convention where $\prod_{j=1}^M \hat{U}_j = \hat{U}_M\ldots \hat{U}_1$. The objective function for this optimization is simply the expectation of the cost Hamiltonian after applying $\hat{U}(\theta)$ to the initial state. This sequence of exponentials is the quantum alternating operator ansatz \cite{farhi2014quantum,hadfield2017quantum}. This is an algorithm which is well-suited for the NISQ era as the number of gates scales linearly with $P$, the exponentials of $\hat{H}_M$ and $\hat{H}_C$ are usually easy to compile without any need for approximation via Trotter-Suzuki decomposition \cite{suzuki1990fractal}. The cost Hamiltonian is typically a sum of terms that are diagonal in the computational basis and often simple to compile. Furthermore, there is no need to split the quantum expectation estimation over multiple runs in order to estimate the various terms; each repetition yields an estimate of the cost function directly.

Now that we have introduced the general QAOA approach, we can explore the specialization of the QAOA to two specific domains of application; namely, MaxCut QAOA and QAOA for Sherrington-Kirkpatrick Ising models. 

\subsubsection{MaxCut QAOA}
The problem for which the QAOA was first explored was for MaxCut \cite{farhi2014quantum}. Let us first provide a brief introduction to the MaxCut problem. Suppose we have a graph $\mathcal{G}= \{\mathcal{V},\mathcal{E}\}$ where $\mathcal{E}$ are the edges and $\mathcal{V}$ the vertices. Given a partition of these vertices into a subset $\mathcal{P}_0$ and its complement $\mathcal{P}_1 = \mathcal{V}\setminus \mathcal{P}_0$, the corresponding \textit{cut set} $C\subseteq \mathcal{E}$ is the subset of edges that have one endpoint in $\mathcal{P}_0$ and the other endpoint in $\mathcal{P}_1$. The \textit{maximum cut} (MaxCut) for a given graph $\mathcal{G}$ is the choice of $\mathcal{P}_0$ and $\mathcal{P}_1$ which yields the largest possible cut set. The difficulty of finding this partition is well known to be an NP-Complete problem in general.

To translate this problem to a quantum Hamiltonian, we can assign a qubit to each vertex $j\in\mathcal{V}$. The computational basis states of these qubits can then be used as binary labels to indicate which partition each qubit is in, i.e., if the qubit $j$ is in the state $\ket{l}_j$, $l \in \{0,1\}$, we assign it to the partition $\mathcal{P}_l$. We can evaluate the size of a cut by counting how many edges have endpoints in different partitions. In order to do this counting, we can compute the XOR of the bit values for the endpoints of each edge and add up these clauses. This cut cardinality can thus be encoded into the cost Hamiltonian for the QAOA as follows: 
\begin{equation}\label{eq:maxcut_ham}
    \hat{H}_{C} = \sum_{\{j,k\}\in{\mathcal{E}}} \tfrac{1}{2}(\hat{I}- \hat{Z}_j \hat{Z}_k).
\end{equation}
Now, for our choice of mixer, the standard choice is the sum of Pauli $\hat{X}$ on each qubit, $\hat{H}_M = \sum_{j\in\mathcal{V}} \hat{X}_j$. This is a good choice as each term is non-commuting with the cost Hamiltonian and it is easy to exponentiate with minimal gate depth. The standard choice of initial state is the uniform superposition over computational bitstrings $\ket{+}^{\otimes |\mathcal{V}|}$, which is an eigenstate of the mixer Hamiltonian. We can now construct our ansatz following \eqref{eq:QAOA} by choosing some value for $P$ and substituting in our MaxCut $\hat{H}_C$ and $\hat{H}_M$. By applying and variationally optimizing the QAOA, one obtains a wavefunction which, when measured in the computational basis, has a high probability of yielding a bitstring corresponding to a partition of large cut size \cite{farhi2014quantum}.

In order to train and test the RNN optimizer on MaxCut QAOA problems, we generated random problem instances in the following fashion: we first fixed an integer $n$, and then randomly sampled an integer uniformly from the range $k \in [3, n - 1]$. Finally, we tossed a random graph from $\mathcal{G}_{n,p}$ with $p=k/n$ and constructed the corresponding MaxCut QAOA QNN of the form of \eqref{eq:QAOA} for $P=2$. Note that a random $\mathcal{G}_{n,p}$ graph is a graph on $n$ nodes where an edge between any two nodes is added independently with probability $p$. To generate training data, we uniformly sampled $n \in [6,9]$, yielding QNN system sizes of at most nine qubits. To train the RNN, 10000 sampled instances from this training set were used. To generate our testing data, we fixed $n=12$, yielding QNN system sizes of 12 qubits, and sampled 50 instances using the procedure described above. 

It has been observed that for random 3-regular graphs, at fixed parameter values of the QAOA ansatz, the expected value of the cost function $\braket{\hat{H}_C}_{\bm{\theta}}$ concentrates \cite{brandao2018fixed}. Our results displayed in Figures \ref{fig:results} and \ref{fig:concentration} corroborate this finding while operating on a slightly broader ensemble of random graphs. This is made clear by noting that initially the MaxCut QAOA has a much narrower 95\% confidence interval across problem instances regardless of optimization algorithm when compared to Ising (SK) QAOA.

\subsubsection{Ising QAOA}

Another domain of application where we tested quantum-classical meta-learning was with the QAOA for finding low energy states of a type of Ising spin glass model known as the Sherrington-Kirkpatrick (SK) model. Many problems in combinatorial optimization can be mapped to these models \cite{kadowaki1998quantum} 
(for example, training Boltzmann machine neural networks \cite{amin2018quantum,verdon2017quantum}). 
In general, finding the lowest energy state of such models is known to be NP-Hard. Using the QAOA, we aim to find low-energy states of an SK Ising Hamiltonian on the graph $\mathcal{G} =\{\mathcal{V},\mathcal{E}\}$, which has the form
\begin{equation}
\hat{H}_C= \tfrac{1}{\sqrt{n}}\!\!\!\sum_{\{j,k\}\in\mathcal{E}}\!\!\! J_{jk}\hat{Z}_j\hat{Z}_k+\sum_{j\in\mathcal{V}} h_j \hat{Z}_j
\label{eq:SK_ham}
\end{equation}
where $n= |\mathcal{V}|$ is the number of vertices, and $J_{jk}$ and $h_j$ are coupling and bias coefficients. For our numerical experiments we considered only the case of the fully connected model where $\mathcal{G}$ is the complete graph. Like the MaxCut QAOA, the choice of mixer Hamiltonian is the sum of the transverse field on all the qubits, $\hat{H}_M = \sum_{j} \hat{X}_j$, and the initial state is chosen as a uniform superposition over all computational basis states $\ket{+}^{\otimes n}$. The parametric ansatz is once again in the form of a regular QAOA (as in \eqref{eq:QAOA}), now with the SK Ising Hamiltonian \eqref{eq:SK_ham} as the cost Hamiltonian. In similar fashion to MaxCut, when we optimize the parameters for this QAOA, we  obtain a wavefunction which, when measured in the computational basis, will yield a bit string corresponding to a spin configuration with a relatively low energy.

Let us now outline the methods used to generate the training and testing data for the RNN specializing in the optimization of Ising QAOA ansatz parameters. To generate random instances of Ising QAOA, we sampled random values of $J_{jk}$, $h_j$ and $n$. For both the training and testing data, after drawing a value for $n$, the parameters $J_{ij}$ and $h_i$ were drawn from independent Gaussian distributions with zero mean and unit variance. Finally, we constructed the corresponding Ising QAOA QNN ansatze of the form \eqref{eq:QAOA} with $P=3$ for the sampled Hamiltonian. For the training instances, we sampled the number of qubits uniformly from $n \in [6,8]$, yielding QNN system sizes of at most 8 qubits. The size of the training set was of 10000 instances from the above described distribution. For testing, we drew 50 samples uniformly from $n \in [9,11]$, thus testing was done with strictly larger instances than those contained in the training set. 

\subsection{Variational Quantum Eigensolvers}
\label{sec:VQE}
\subsubsection{Hubbard Model VQE}

Here we describe the variational quantum eigensolver (VQE) ansatze that were used to generate the results in Fig. \ref{fig:results}. The specific class of VQE problems we chose to consider were for variational preparation of ground states of Hubbard model lattices \cite{kivlichan2018quantum}. The Hubbard model is an idealized model of fermions interacting on a lattice. The 2D Hubbard model has a Hamiltonian of the form $\hat{H} = \hat{T}_h + \hat{T}_v + \hat{V}$, where $\hat{T}_h$ and $\hat{T}_v$ are the horizontal and vertical hopping terms and $\hat{V}$ a spin interaction term, more explicitly,
\begin{equation}\label{eq:Hubbard}
    \hat{H}
    = -t \!\!\sum_{\langle i, j \rangle, \sigma}\!\! (\hat{a}_{i, \sigma}^\dagger \hat{a}_{j, \sigma} + \hat{a}_{j, \sigma}^\dagger \hat{a}_{i, \sigma})
    + U \sum_{i} \hat{a}_{i, \uparrow}^\dagger \hat{a}_{i, \uparrow} \hat{a}_{i, \downarrow}^\dagger \hat{a}_{i, \downarrow} \\
\end{equation}
where the $\hat{a}_{j, \sigma}$ and $\hat{a}_{j, \sigma}^\dagger$ are annihilation and creation operators on site $j$ with spin $\sigma \in \{\uparrow,\downarrow\}$. The goal of the VQE is to variationally learn a parametrized circuit which prepares the ground state of the Hubbard Hamiltonian from \eqref{eq:Hubbard}, or at least an approximation thereof.

Our variational ansatz to prepare these approximate ground states is based on the Trotterization of the time evolution under the Hubbard model Hamiltonian, it is of the form
\begin{equation}
\hat{U}(\bm{\theta}) = \prod_{j=1}^Pe^{-i \theta^{(j)}_h \hat{T}_h}  e^{-i\theta^{(j)}_v \hat{T}_v}  e^{-i\theta^{(j)}_U \hat{V}}
\end{equation}
where $\bm{\theta}= \{\bm{\theta}_h,\bm{\theta}_v,\bm{\theta}_U\}$ are the variational parameters for the $P$ Trotter steps. The exponentials at each step are done using a single fermionic swap network \cite{kivlichan2018quantum}.   This is similar to the ansatz used in \cite{wecker2015progress} but corresponds to a different order of simulation of the terms.

Let us provide more details as to our choices of parameters used to generate the results from Figure \ref{fig:results}.
We used an ansatz consisting of $P=5$ steps, where each step introduced 3 parameters.  For our initial state, we use an eigenstate of the kinetic term with the correct particle number and the same total spin as the ground state, and we study the model at half-filling. We set $t = 1.0$ for all instances, and this defines our units of energy. Our training data consists of 10000 instances with the lattice system size chosen to be either $n=2 \times 2$ or $n=3 \times 2$ with equal probability and with $U$ chosen from a uniform distribution on the domain of $[0.1, 4.0]$. After training, we tested the neural network on instances with system size $n=4 \times 2$, again strictly larger than our training set.

\subsection{Meta-learning Methods \& Results}
\label{sec:Results}

\begin{figure*}\label{fig:results}

    \text{\hspace{30pt} MaxCut QAOA \hspace{100pt} Ising QAOA \hspace{90pt}  VQE Hubbard Model}

    \includegraphics[width=0.32\textwidth,trim={0.48cm 0.40cm 0.48cm 0.32cm},clip]{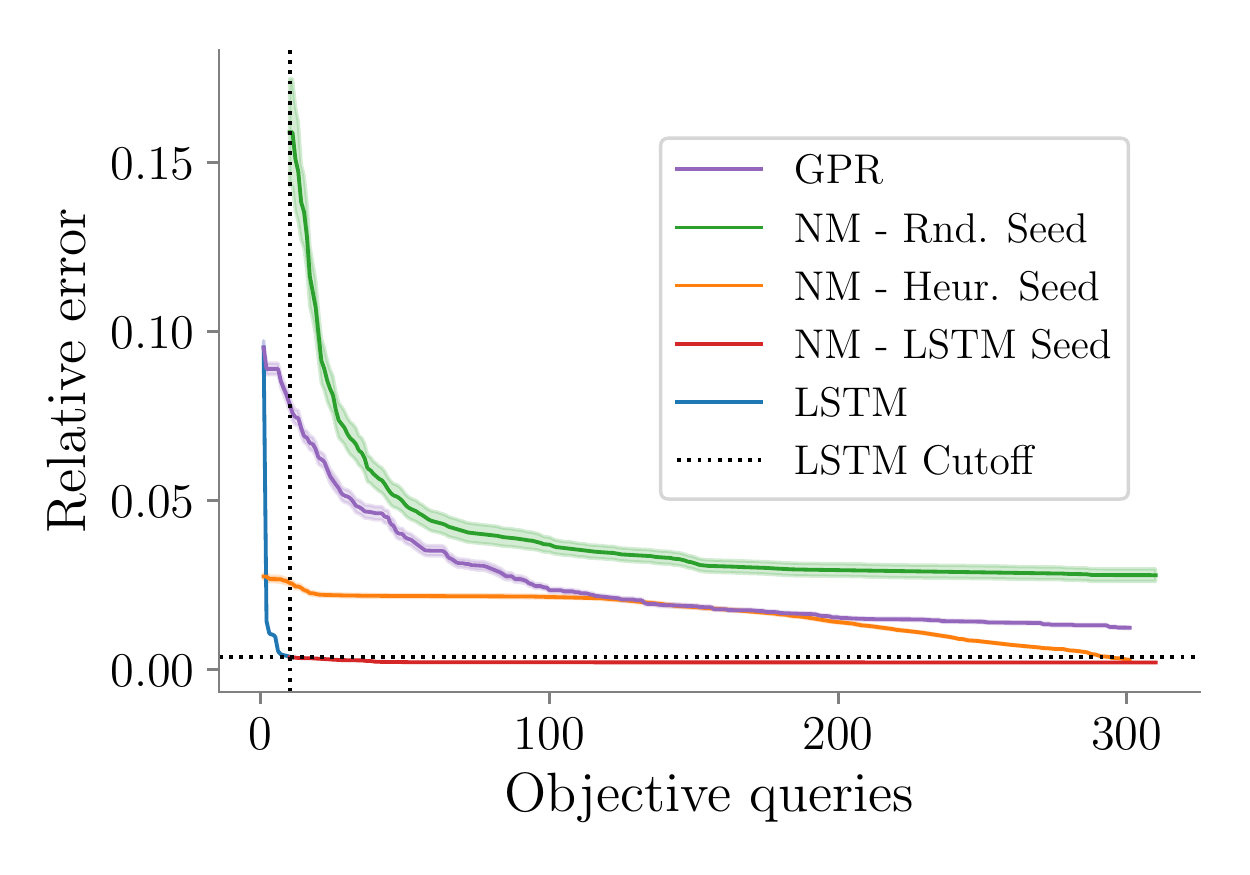}
    \includegraphics[width=0.32\textwidth,trim={0.48cm 0.40cm 0.48cm 0.32cm},clip]{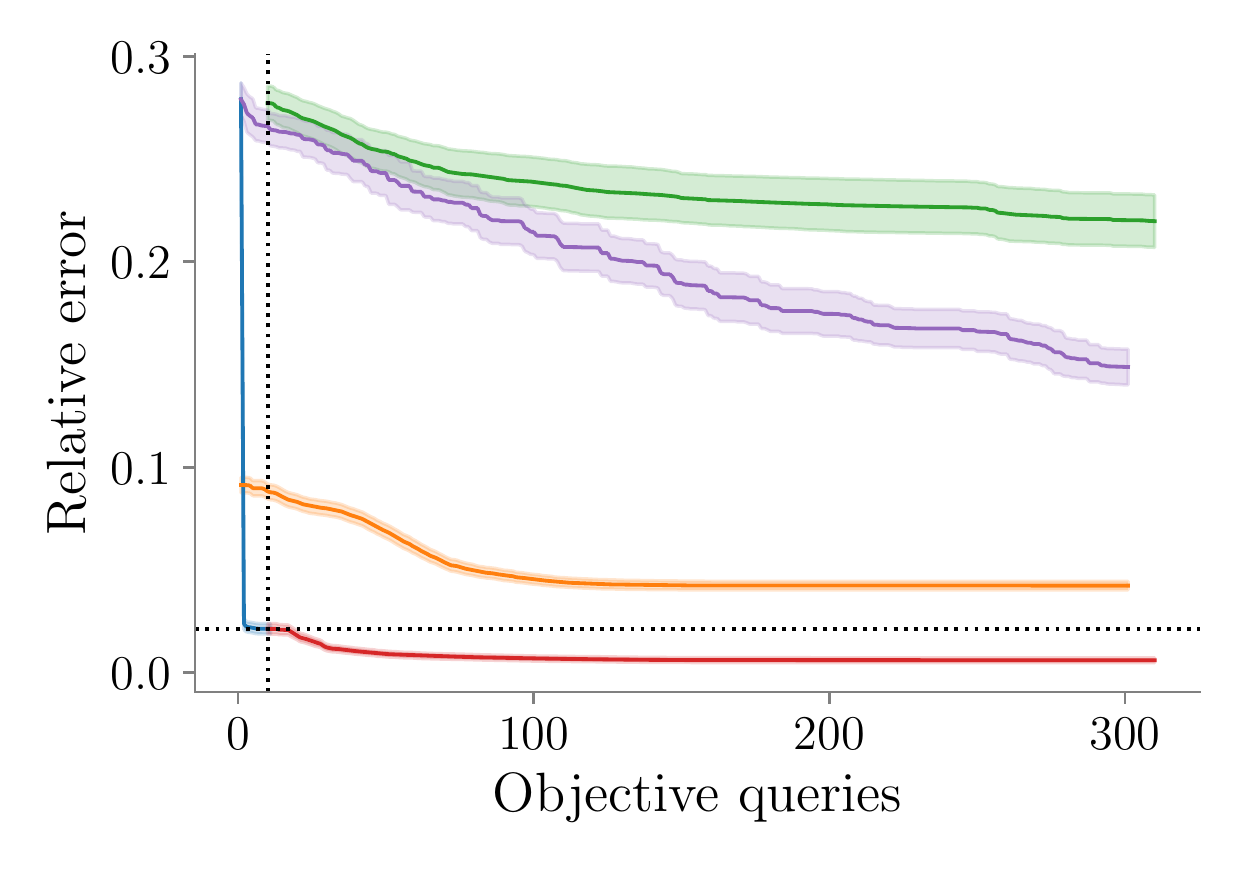}
    \includegraphics[width=0.32\textwidth,trim={0.48cm 0.40cm 0.48cm 0.32cm},clip]{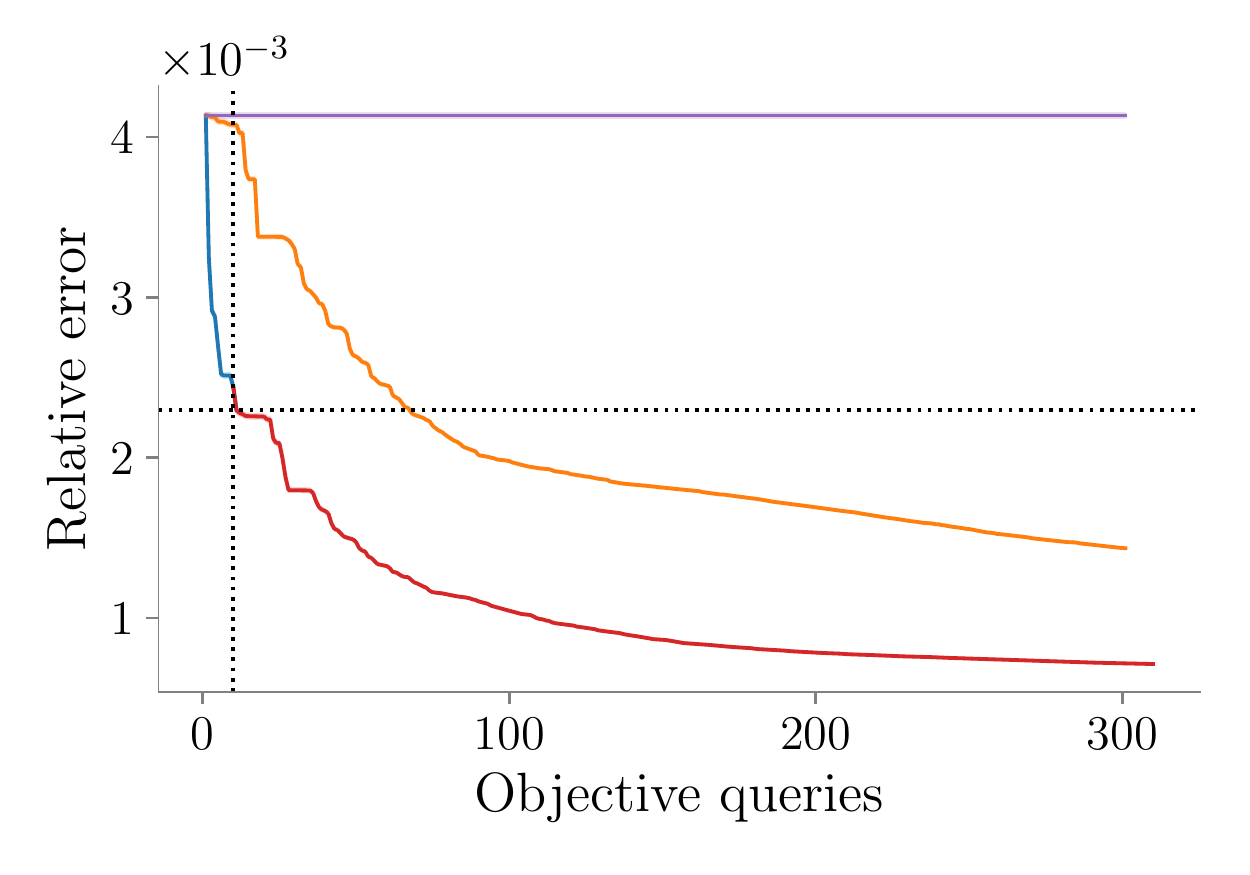}
    
    \includegraphics[width=0.32\textwidth,trim={0.48cm 0.40cm 0.48cm 0.32cm},clip]{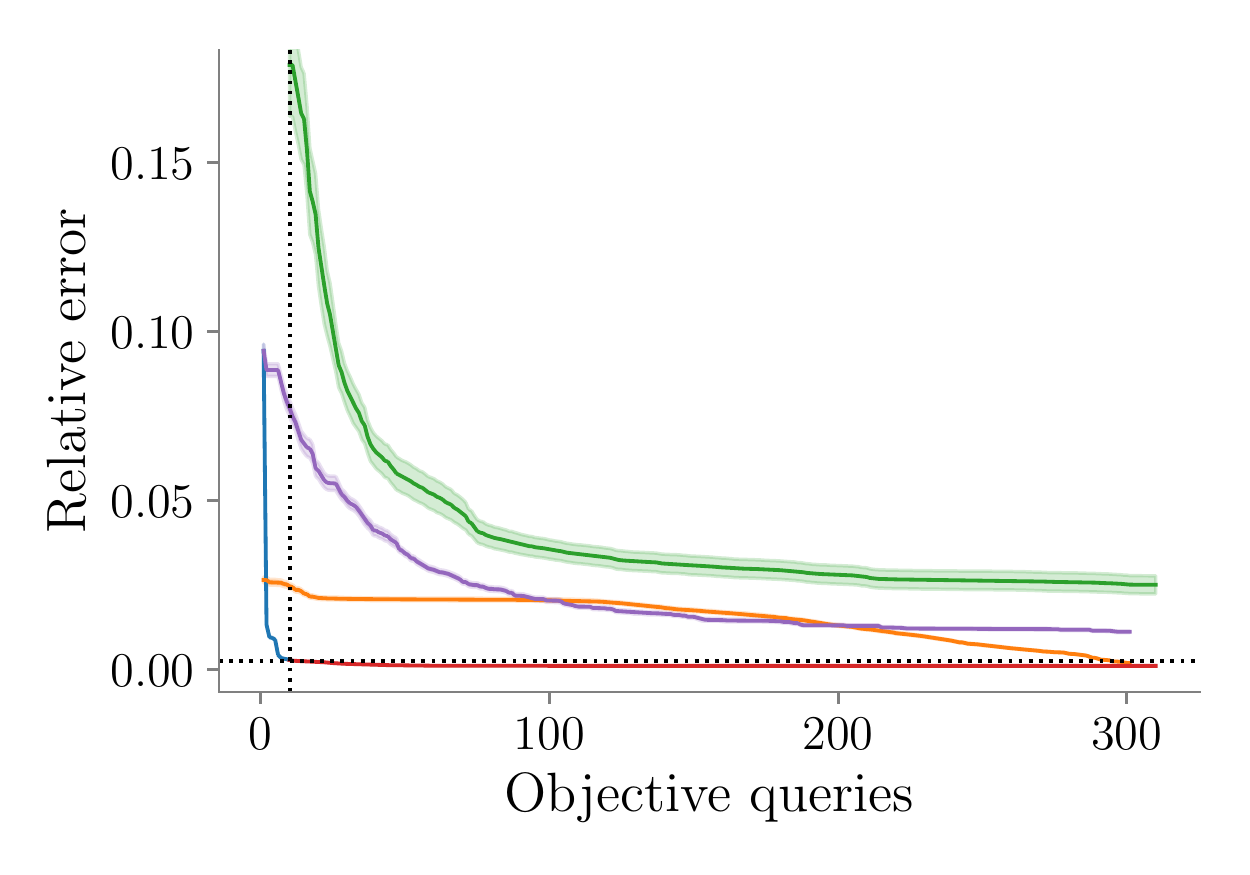}
    \includegraphics[width=0.32\textwidth,trim={0.48cm 0.40cm 0.48cm 0.32cm},clip]{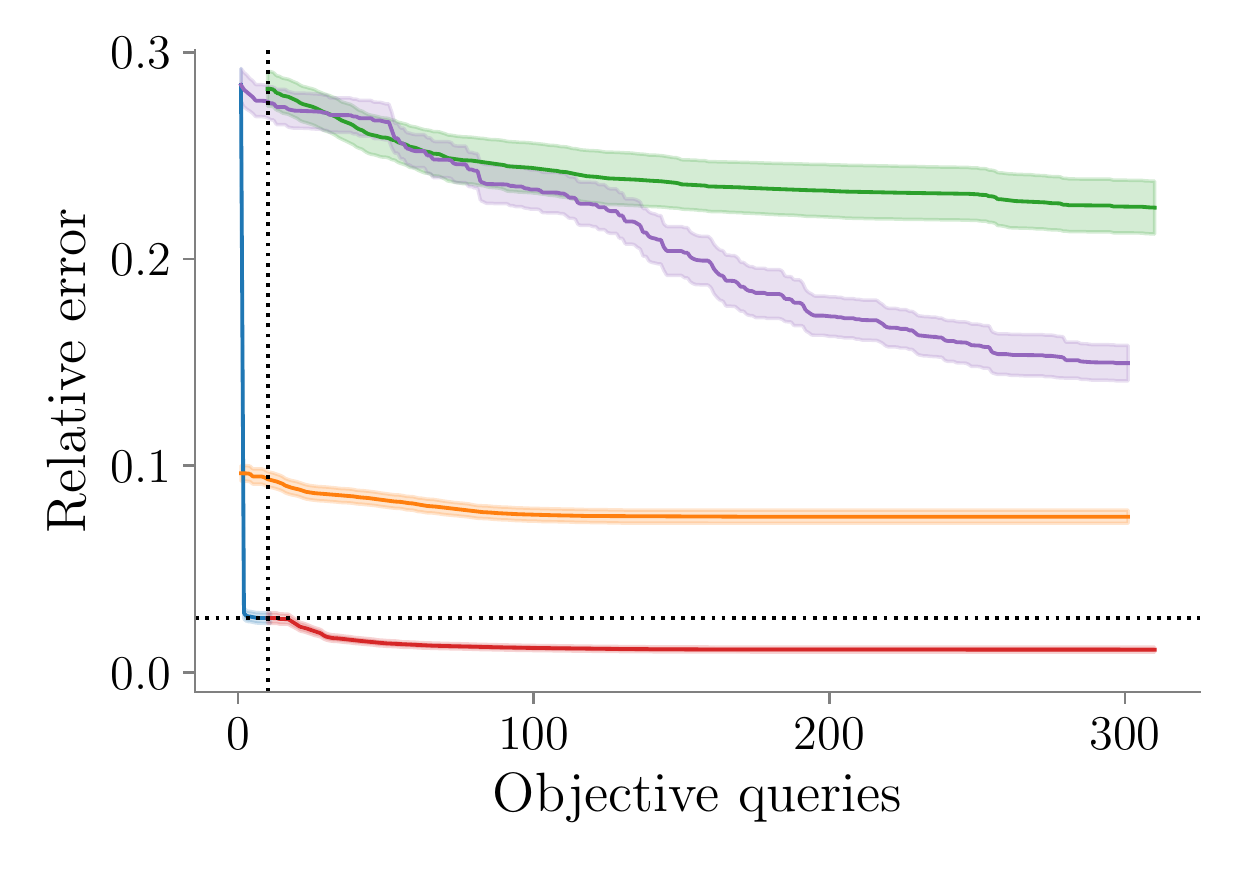}
    \includegraphics[width=0.32\textwidth,trim={0.85cm 0.40cm 0.48cm 0.32cm},clip]{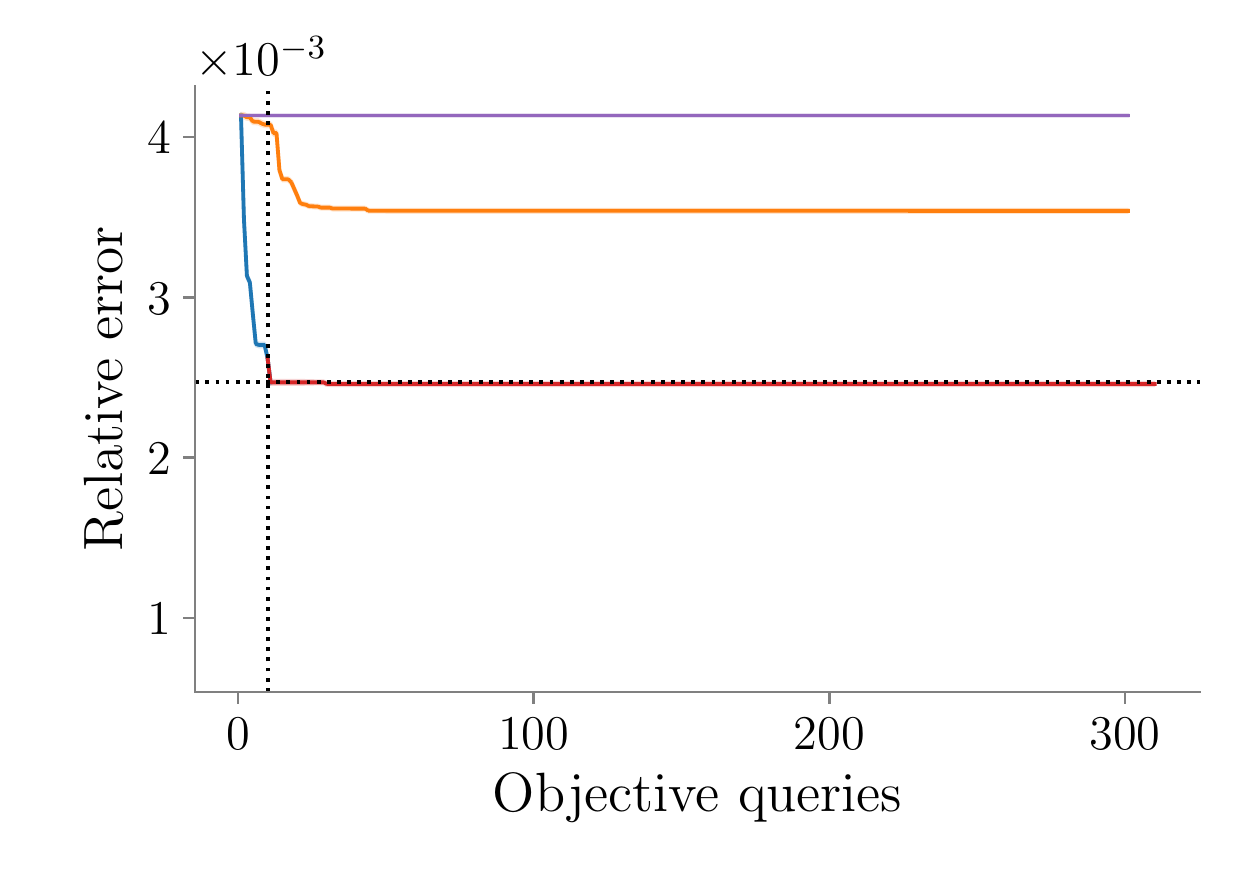}

    \caption{Displayed above are the average relative errors with respect to the number of objective function queries during the training of 50 random problem instances for the three classes of problems of interest, QAOA for MaxCut (left), QAOA for Ising models (middle), and VQE for the Hubbard model (right), for various choices of optimizers and initialization heuristics. The problem instances were sampled from the testing distribution described in sections \ref{sec:QAOA} and \ref{sec:VQE}. These include a Gaussian Process Regression (GPR) optimizer \cite{rasmussen2004gaussian}, and Nelder-Mead (NM) \cite{lagarias1998convergence} with various initialization heuristics. The first of these initialization heuristics was the best of 10 random guesses seed (Rnd. Seed). Also presented is NM initialized with application-specific heuristic seeds (Heur. Seed), which consisted of the adiabatic heuristic for VQE \cite{mcclean2017openfermion}, and the mean optimal parameters of the training set for QAOA \cite{brandao2018fixed}, and finally the seed from our meta-learned neural optimizer (LSTM). We cut off the LSTM after 10 iterations, as it is used mainly as an initializer for other optimizers. Note that we have not included overhead of the meta-training in this plot, see the main text for a breakdown of the overhead for the training of the LSTM. The top row is for noiseless readout of the expectation, while the bottom row has some Gaussian noise with variance 0.05 added to the expectation value readouts, thus emulating approximate estimates of expectation values. For reference, for the set of testing instances, given a QPU inference repetition rate of 10 kHz, the necessary wall clock time per objective query to achieve this variance \cite{wecker2015progress} in the cost estimate is at most $(70\pm40)$ seconds for MaxCut QAOA, $(2.3 \pm 0.6)$ seconds for Ising QAOA, and $(30 \pm 20)$ seconds for Hubbard VQE. Note that the relative error is the difference in the squashed cost function relative to the squashed global optimum found through brute force methods, i.e., $\bar{f}_{\text{rel}}((\bm{\theta}))=(\bar{f}(\bm{\theta})-\min\bar{f})$.  Error bars represent the 95\% confidence interval for the random testing instances from the distribution of problems described in sections \ref{sec:QAOA} and \ref{sec:VQE}.}\label{fig:results}

\end{figure*}

In this section we present the main results of our quantum-classical meta-learning experiments, displayed in Figure \ref{fig:results}, and discuss some additional details of our methods used to produce these results. We trained and tested a set of long short-term memory (LSTM) recurrent neural networks (RNN) to learn to optimize the variety of QNN instances discussed in sections \ref{sec:QAOA} and \ref{sec:VQE}, namely, MaxCut QAOA, Ising QAOA and Hubbard VQE. For each of the three problem classes, the RNN was trained using 10000 problem instances. This training of the RNN was executed over a maximum of 1000 epochs, each with a time horizon of 10 iterations. Hence, training required the simulation of inference for at most 1 million quantum neural networks. In most cases the meta-training was stopped well before these 1000 epochs were completed, following standard early-stopping criteria \cite{prechelt1998early}.

The quantum circuits used for training and testing the recurrent neural network were executed using the Cirq quantum circuit simulator \cite{Cirq} running on a classical computer. The VQE ansatze were built using OpenFermion-Cirq \cite{mcclean2017openfermion}. Neural network training and inference was done in TensorFlow \cite{abadi2016tensorflow}, using code adapted from previous work by DeepMind \cite{chen2016learning}.

For both testing and training, we squashed the readout of the cost function by a quantity which bounds the operator norm of the Hamiltonian. This was done to ensure a normalized loss signal for our RNN across various problem instances. In classical machine learning, normalizing data variance is well-known to accelerate and ameliorate training \cite{ioffe2015batch}. In the same spirit, we fed the RNN a cost function squashed according to the Pauli coefficient norm, denoted $\lVert \ldots\rVert_*$. Recall that for a Hamiltonian $\hat{H}$ with a decomposition as a linear combination of Paulis of the form $\hat{H} = \sum_{j}\alpha_j \hat{P}_j$, then  $\lVert \hat{H}\rVert_* \equiv \lVert\alpha\rVert_1 =  \sum_j |\alpha_j|$. The squashed cost function is then simply the regular expectation value of the Hamiltonian, divided by the Pauli coefficient norm, $
    \bar{f}(\bm{\theta}) = \braket{\hat{H}}_{\bm{\theta}}/\lVert\hat{H}\rVert_*.$
As all Paulis have a spectrum of $\{\pm1\}$ we are guaranteed that the squashed cost function $ \bar{f}(\bm{\theta})$ has its range contained in $[-1,1]$. In Figure \ref{fig:results}, we plot the \textit{relative error}, which is the difference in the squashed cost function relative to the globally optimal squashed cost function value found through brute force methods, $\bar{f}_{\text{rel}}(\bm{\theta})=(\bar{f}(\bm{\theta})-\min_{\bm{\theta}}\bar{f}(\bm{\theta}))$. The brute force optimization methods were basin hopping for the QAOA instances \cite{wales1997global}, and exact diagonalization for the VQE instances.

For the testing of the trained LSTM, we used randomly sampled instances from the distributions of ansatze described in sections \ref{sec:QAOA} and \ref{sec:VQE}. In all cases, the testing instances were for larger-size systems than those used for training, while keeping the number of variational parameters of the ansatze fixed. Note that as all the ansatze considered in this paper were QAOA-like, one can thus scale the size of the system while keeping the same number of parameters. This is an important feature of this class of ansatze as our LSTM is trained to optimize ansatze of a fixed parameter space dimension. 

For all instances, the LSTM was trained on a time horizon of $T=10$ quantum-classical iterations, using the observed improvement \eqref{eq:OI} as the meta-learning loss function. We trained the LSTM on noiseless quantum circuit simulations in Cirq \cite{Cirq}. Note that training of each of the three LSTM networks already required the simulation of 1 million quantum circuit executions with the chosen time horizon of 10 iterations, and that the number of quantum circuit simulations scales linearly with the time horizon. Additionally, gradient-based training required backpropagation through time for the temporal hybrid quantum-classical computational graph, which added further linearly-scaling overhead. Thus, we chose a short time horizon to minimize the complexity of the training. For reference, 10 iterations is a significantly smaller number of quantum-classical optimization iterations than what is typically seen in previous works on QNN optimization \cite{nannicini2018performance}. The typical number of iterations required by other optimizers is usually on the order of hundreds to possibly thousands to reach a comparable optimum of the parameter landscape.

 Although the LSTM reaches a good approximate optimum in these 10 iterations, some applications of QNN's such as VQE require further optimization as a high-precision estimate of the cost function is desired. Thus, instead of simply using the LSTM as an optimizer for an extended time horizon, we used the LSTM as a few-iteration initializer for Nelder-Mead (NM). This was done to minimize the complexity of training the RNN and avoid the instabilities of longer training horizons where the RNN would most likely learn a local method for fine-tuning its own initial guess. A longer time horizon would thus most likely not have provided a significant gain in performance, all the while substantially increasing cost of training.

We tested the robustness of the RNN optimizer by comparing its performance to other common optimizers, both in the cases where Gaussian noise was added to the cost function evaluations, and for a noiseless readout idealized case. This additional Gaussian noise can be interpreted as a means to emulate the natural noise of quantum expectation estimation with a finite number of measurement runs \cite{mcclean2016theory}. Figure \ref{fig:results} allows for comparison of noisy and noiseless inference (QNN optimization) for the trained LSTM versus alternative optimization and intilization heuristics. For the noisy tests, the expectation samples obeyed a normal distribution of variance 0.05, thus the cost function estimates were drawn according to $y_t \sim \mathcal{N}(\braket{\hat{H}}_{\bm{\theta}_t}, 0.05)$ for the results presented in Figure \ref{fig:results}. For the testing instances used to generate the results presented in Figure \ref{fig:results}, following a standard prescription \cite{wecker2015progress} for the number of repetitions required to guarantee an upper bound to the variance of 0.05, the number of repetitions (QNN inference runs) should be of $(7\pm4)\times 10^5$ repetitions for MaxCut QAOA, $(2.3 \pm 0.6)\times 10^4$ repetitions for Ising QAOA, and $(3 \pm 2)\times 10^5$ repetitions for Hubbard VQE. In terms of wall clock time, assuming that the QPU can execute 10000 repetitions (consisting of a quantum circuit execution, multi-qubit measurement, and qubit resetting) per second, for the distribution of testing instances, the total time needed for the LSTM to perform its 10 optimization steps is in the range of $(700\pm400)$ seconds for MaxCut QAOA, $(23 \pm 6)$ seconds for Ising QAOA, and $(300 \pm 200)$ seconds for Hubbard VQE. Note that the standard deviation here is due to the variations in the Pauli norm of the Hamiltonians for the sampled instances of the test set.

Apart from this added cost function noise, the simulated quantum circuit executions were simulated without any other form of readout or gate execution noise. Plotted in Figure \ref{fig:results} are the $95\%$ confidence intervals for the optimization of the 50 testing instances which were sampled according to the testing distributions described in Sec. \ref{sec:QAOA} and Sec. \ref{sec:VQE}. Our results show that the neural optimizer learns initialization heuristics for the QAOA and VQE parameters which generalize across problem sizes. We discuss these results in further detail in the following section.

Let us provide a description of the alternative optimization and initialization heuristics used to generate Figure \ref{fig:results}. First alternative strategy was a Bayesian Optimization using Gaussian processes \cite{rasmussen2004gaussian}, here the initial parameters are set to nil, same as the was the case for the LSTM optimizer. Next, in order to compare the LSTM to other initialization heuristics, we compared the initialization of Nelder-Mead (NM) at parameter values found from the best of 10 random guesses (Random Seed), NM initialized using some state of the art heuristics for QAOA and VQE (Heuristic Seed), and NM initialized after 10 iterations of the LSTM (LSTM Seed). The  application-specific heuristic seeds (Heur. Seed) were the adiabatic heuristic for VQE \cite{mcclean2017openfermion}, where the variational parameters are scaled in a similar fashion to an adiabatic interpolation across the 5 steps, while for the QAOA the parameters were initialized at the mean value of the optimal parameters for the training set distribution of problem instances. As was shown in \cite{brandao2018fixed}, as there is a concentration of the cost function for fixed parameters, one can thus expect the distribution of optimal parameters of the QAOA to be concentrated around some mean. 

\begin{figure}\label{fig:concentration}

    \footnotesize{\text{\hspace{15pt} MaxCut QAOA\hspace{60pt} Ising QAOA}\hspace{20pt}}

    \includegraphics[width=0.5\columnwidth]{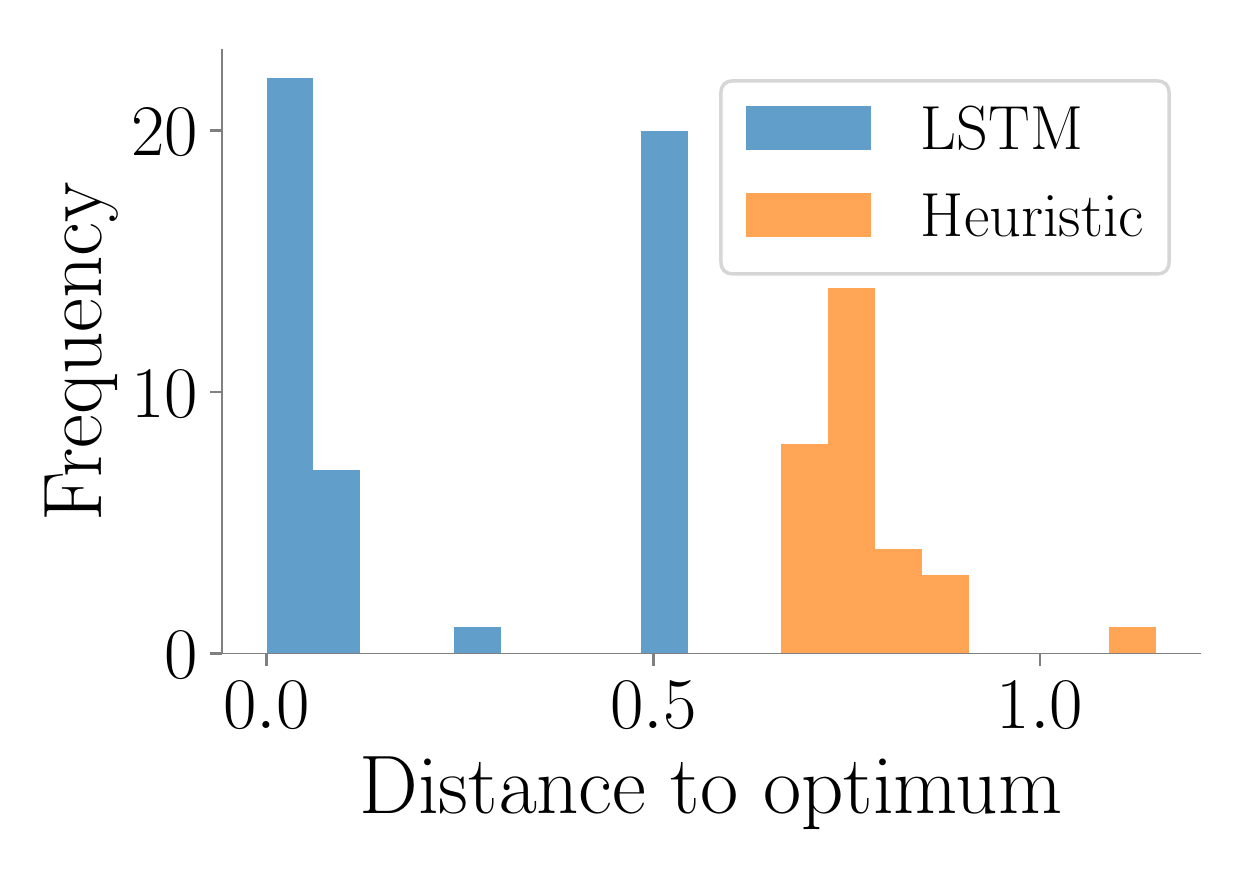}
    \includegraphics[width=0.5\columnwidth]{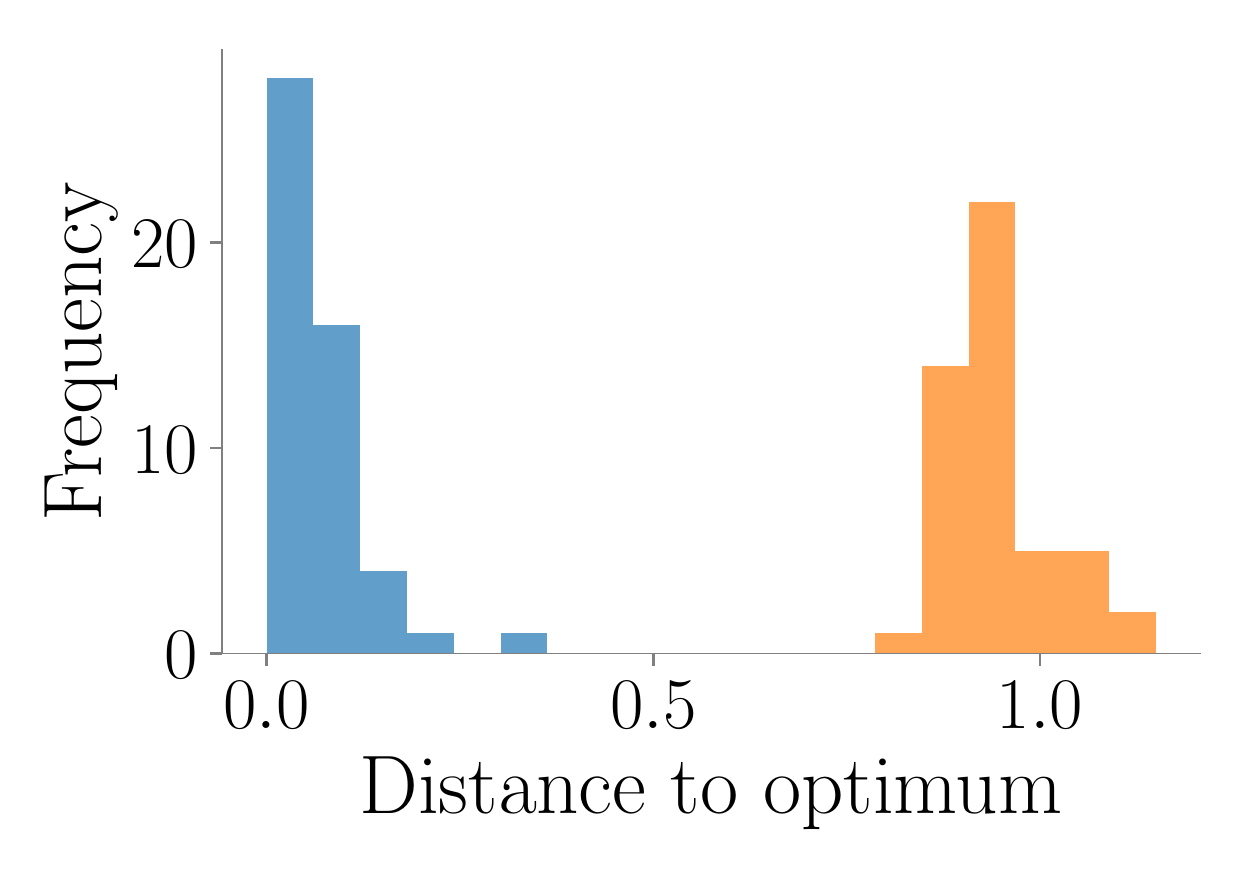}
    
    \footnotesize{\text{\hspace{15pt} VQE Hubbard Model}}
    
    \includegraphics[width=0.5\columnwidth,
    ]{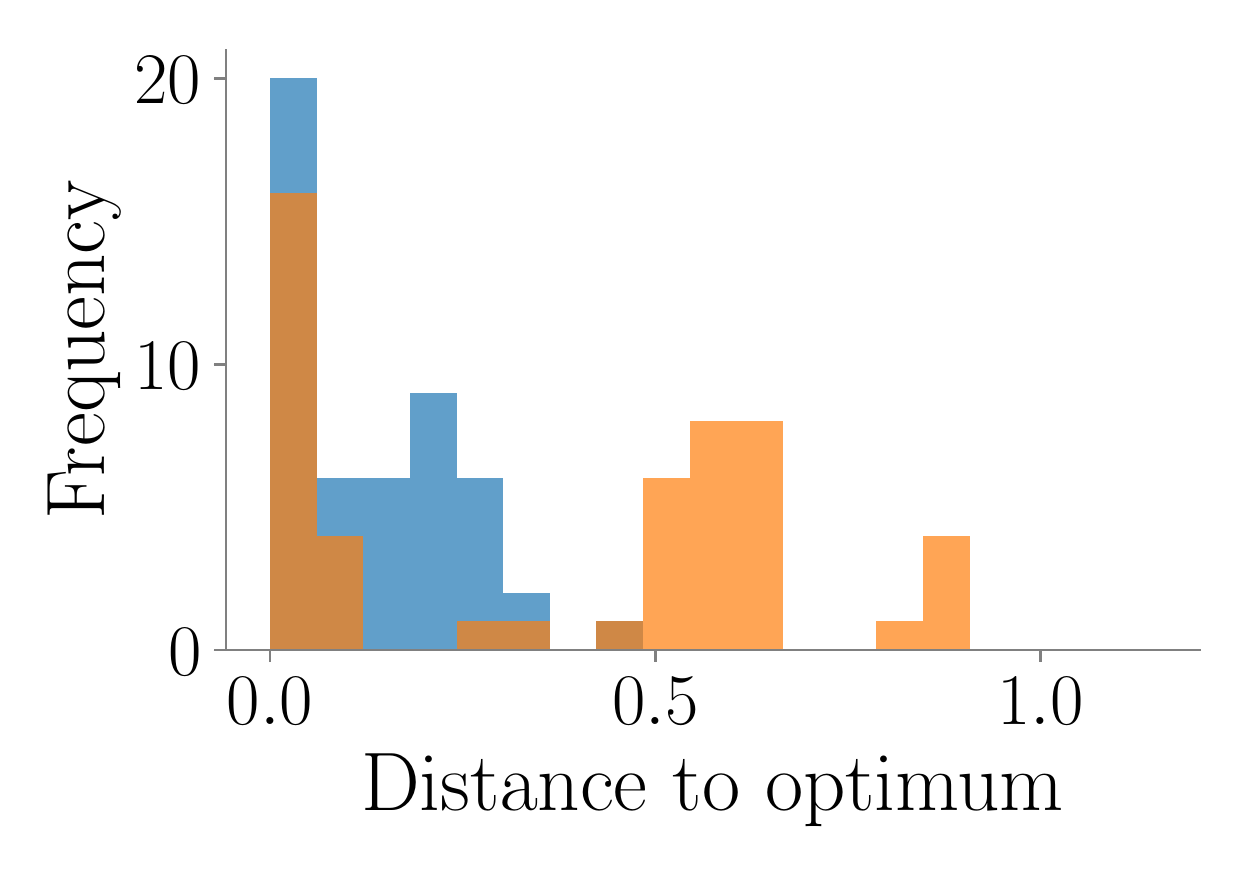}

    \caption{The above histograms represent the parameter space Euclidean distance to the true optimum $d(\bm{\theta})= \lVert \bm{\theta}-\bm{\theta}^*\rVert_2$, immediately after initialization, for both the LSTM (after 10 iterations) and alternative initialization heuristics. For each of the three cases, 50 samples of the test set were used. For the QAOA, this alternative initialization heuristic was identical to \cite{brandao2018fixed}, whereas for VQE, this was the Adiabatic Heuristic \cite{mcclean2017openfermion}. The histograms were collected from the test set problem instances, which were described in Sections \ref{sec:QAOA} and \ref{sec:VQE}, and were used to generate the results presented in Figure \ref{fig:results}. We see that the LSTM initializes parameters significantly closer to the optimum on average as compared alternative heuristics.
    }
\end{figure}

In Figure \ref{fig:concentration}, we compare the Euclidean distance in parameter space between the output of the 10 iterations of the LSTM versus other initialization heuristics. We clearly see that the LSTM optimizer initializes the QNN parameters closer to the optimal parameters of each test instance, on average, as compared to other methods. We see that in the case of the QAOA, the constant fit heuristic \cite{brandao2018fixed} for the training instances yields a cluster of parameters that is not clustered around the optimal parameters of the larger instance, while the LSTM output parameters are significantly closer to the globally optimal parameters found by brute force. This shows a clear separation between the parameters obtained from a constant fit of the training set versus the LSTM's adaptive scheme for optimizing parameters in few iterations.


\section{Discussion}

As shown in Figure \ref{fig:results}, our trained neural optimizer reaches a higher-quality approximate optimum of the parameters in 10 iterations than other optimizers can manage in hundreds, both for noisy and noiseless readout. Most evident in the case of VQE, where the local optimizers can have severe difficulty optimizing parameters when given noisy evaluations of the cost function. Of all alternatives to the neural optimizer, the probabilistic approach of Bayesian optimization via Gaussian process regression was the best performer.

In all six settings, the LSTM rapidly finds an approximate minimum within its restricted time horizon of 10 iterations. The neural optimizer needs to initialize the parameters in a basin of attraction of the cost function landscape so that a local optimizer can then easily converge to a local optimum in fewer iterations and more consistently. As we can see across all cases, the Nelder-Mead runs that are initialized by the LSTM rather than other initialization heuristics tend to reach the highest quality optimum with much lower variance in performance. These result show that the LSTM initializes the parameters in a good basin of attraction near the optimum. Although the LSTM initialization helps in all cases, we can see that for the noisy VQE case in Figure \ref{fig:results}, the Nelder-Mead approach struggles to improve upon the guess of the LSTM. While not the focus of this work, these results point towards the need for further investigation into better local optimization techniques which are robust to noise \cite{o2016scalable}.

When looking at the optimal parameters found by the RNN (see Fig. \ref{fig:concentration}), we observed a certain degree of concentration in their values, similar to what has been observed in previous works \cite{brandao2018fixed,yang2017optimizing}. The MaxCut QAOA was the most concentrated, which corroborates recent observations \cite{brandao2018fixed}. Similarly, the optimal parameters of the Ising QAOA were also observed to have a degree of concentration, as was also observed in previous work \cite{yang2017optimizing}. Finally, the VQE had the least amount of concentration of the three problem classes, but still exhibited some degree of clustering in the optimal values. 

The concentration of parameters is not surprising given the connections between QAOA-like ansatze and gradient descent/adiabatic optimization \cite{verdon2018universal,bapat2018bang}. 
In a sense, these QAOA-like QNN's are simply variational methods to descend the energy landscape, and the variational parameters can be seen as energetic landscape descent hyperparameters, akin to gradient descent parameters. Similar to how the classical meta-learning approaches to gradient-based optimization converged onto methods comparable to best-practices for hyperparameter optimization (e.g., comparable to performance of AdaGrad and other machine learning best-practice heuristics), the neural optimizer in our case found a neighborhood of optimal hyperparameters and learned a heuristic to quickly adjust these parameters on a case-by-case basis. 

Although it may seem that this meta-learning method is costly due to its added complexity over regular optimization, one must remember that, for the time being, classical computation is still much cheaper than quantum computation. As the optimization scheme generalizes across system sizes, one may imagine training an LSTM to optimize a certain ansatz for small system sizes by quantum simulation on a classical computer, then using the LSTM to rapidly initialize the parameters for a much larger instance of the same class of problem on a quantum computer. This approach may be well worth the added classical computation, as it can reduce the number of required runs to get an accurate answer on the QPU by an order of magnitude or more. 

\section{Conclusion \& Outlook}

In this paper, we proposed a novel approach to the optimization of quantum neural networks, namely, using meta-learning and a classical neural network optimizer. We tested the performance of this approach on a set of random instances of variational quantum algorithm optimization tasks, which were the Quantum Approximate Optimization applied to MaxCut and Sherrington-Kirkpatrick Ising spin glasses, and a set of Variational Quantum Eigensolver ansatze for preparation of ground states of Hubbard models. 

The neural network was used to rapidly find a global approximate optimum of the parameters, which then served as an initialization point for other local search heuristics. This combination yielded optima of the quantum neural network landscape which were of a higher quality than alternatives could produce with orders of magnitude more quantum-classical optimization iterations. Furthermore, the neural network exhibited generalization capacity across problem sizes, thus opening up the possibility of classical pre-training of the neural optimizer for inference on larger instances requiring quantum processors.

Two significant challenges of quantum neural network optimization in the NISQ era are finding optimization methods that allow for precise fine-tuning of the parameters to hone in on local minima despite the presence of readout noise and to find good initialization heuristics to allow for more consistent convergence of these local optimizers. Given the results presented in this paper, we believe that this first challenge has been mitigated by our quantum-classical meta-learning approach, while the second challenge remains open for future work.

In terms of possible extensions of this work, the meta-learning approach could be further improved in several ways. One such way would be to use more recent advances in meta-learning optimizer neural networks \cite{wichrowska2017learned} which can scale to arbitrary problems and number of parameters. This would extend the capabilities of our current approach to optimizing the parameters of arbitrary QNN's beyond Trotter-based/QAOA-like ansatze with variable numbers of parameters across instances. Another possible extension of this work would be to meta-learn an optimizer for Quantum Dynamical Descent \cite{verdon2018universal}, a quantum generalization of gradient descent which takes the form of a continuous-variable QAOA. As our neural optimizer was tested on various QAOA problems successfully, one could imagine applying it to the optimization of the Quantum Dynamical Descent hyperparameters. The latter could be considered learning to \textit{learn with quantum dynamical descent with classical gradient descent}. This would also be a way to generalize the applicability of our approach to arbitrary QNN optimization tasks. Finally, as a NISQ-oriented alternative to the latter, one could meta-learn to optimize the hyperparameters for the stochastic quantum circuit gradient descent algorithm recently proposed by Harrow et al. \cite{harrow2019low}. We leave the above proposed explorations to future work.

\section{Acknowledgements}
Circuits and neural networks in this paper were implemented using a combination of Cirq \cite{Cirq}, OpenFermion-Cirq \cite{mcclean2017openfermion}, and TensorFlow \cite{abadi2016tensorflow}. The authors would like to thank Yutian Chen and his colleagues from DeepMind for providing code for the neural optimizer \cite{chen2016learning} which was adapted for this work, as well as Edward Farhi, Li Li, and Murphy Niu for their insights, observations, and suggestions. MB and GV would like to thank the team at the  Google AI Quantum lab for the hospitality and support during their respective internships where this work was completed. GV acknowledges funding from NSERC.

\bibliography{lib}

\end{document}